\documentclass[twocolumn,showkeys,showpacs,preprintnumbers,prd,superscriptaddress,nofootinbib]{revtex4-1}
\usepackage{graphicx,color,bm,verbatim,multirow,epsf,epstopdf,amsmath,amsfonts,amssymb}
\usepackage{natbib}
\usepackage[normalem]{ulem}
\usepackage{hyperref}
\hypersetup{colorlinks=true,urlcolor=blue,citecolor=blue,linkcolor=blue,menucolor=blue,anchorcolor=blue,filecolor=blue}

\begin{document}

\title{New tests of dark sector interactions from the full-shape galaxy power spectrum}

\author{Rafael C. Nunes}
\email{rafadcnunes@gmail.com}
\affiliation{Instituto de F\'{i}sica, Universidade Federal do Rio Grande do Sul, 91501-970 Porto Alegre RS, Brazil}
\affiliation{Divis\~{a}o de Astrof\'{i}sica, Instituto Nacional de Pesquisas Espaciais, Avenida dos Astronautas 1758, S\~{a}o Jos\'{e} dos Campos, 12227-010, S\~{a}o Paulo, Brazil}

\author{Sunny Vagnozzi}
\email{sunny.vagnozzi@ast.cam.ac.uk}
\affiliation{Kavli Institute for Cosmology (KICC) and Institute of Astronomy,\\University of Cambridge, Madingley Road, Cambridge CB3 0HA, United Kingdom}

\author{Suresh Kumar}
\email{suresh.math@igu.ac.in}
\affiliation{Department of Mathematics, Indira Gandhi University, Meerpur, Haryana 122502, India}

\author{Eleonora Di Valentino}
\email{e.divalentino@sheffield.ac.uk}
\affiliation{School of Mathematics and Statistics, University of Sheffield, Hounsfield Road, Sheffield S3 7RH, United Kingdom}

\author{Olga Mena}
\email{omena@ific.uv.es}
\affiliation{Instituto de F\'{i}sica Corpuscular (IFIC), University of Valencia-CSIC, E-46980, Valencia, Spain}

\begin{abstract}
\noindent We explore the role of redshift-space galaxy clustering data in constraining non-gravitational interactions between dark energy (DE) and dark matter (DM), for which state-of-the-art limits have so far been obtained from late-time background measurements. We use the joint likelihood for pre-reconstruction full-shape (FS) galaxy power spectrum and post-reconstruction Baryon Acoustic Oscillation (BAO) measurements from the BOSS DR12 sample, alongside Cosmic Microwave Background (CMB) data from \textit{Planck}: from this dataset combination we infer $H_0=68.02^{+0.49}_{-0.60}\,{\rm km}/{\rm s}/{\rm Mpc}$ and the 2$\sigma$ lower limit $\xi>-0.12$, among the strongest limits ever reported on the DM-DE coupling strength $\xi$ for the particular model considered. Contrary to what has been observed for the $\Lambda$CDM model and simple extensions thereof, we find that the CMB+FS combination returns tighter constraints compared to the CMB+BAO one, suggesting that there is valuable additional information contained in the broadband of the power spectrum. We test this finding by running additional CMB-free analyses and removing sound horizon information, and discuss the important role of the equality scale in setting constraints on DM-DE interactions. Our results reinforce the critical role played by redshift-space galaxy clustering measurements in the epoch of precision cosmology, particularly in relation to tests of non-minimal dark sector extensions of the $\Lambda$CDM model.
\end{abstract}

\keywords{}

\pacs{}

\maketitle

\section{Introduction}
\label{sec:introduction}

Dark energy (DE) accounts for approximately $70\%$ of the Universe's energy budget, yet its nature remains puzzling. Within the standard six-parameter $\Lambda$CDM cosmological model, DE takes the form of a smooth, time-independent, spatially uniform vacuum energy component. This simple model is able to provide an extremely good fit to a wide variety of cosmological and astrophysical observations, including anisotropies in the Cosmic Microwave Background (CMB), the clustering of the large-scale structure (LSS), the magnitude-redshift relation of Type Ia Supernovae (SNeIa) in the Hubble flow, the distortion of images of distant galaxies due to weak lensing from the intervening LSS (cosmic shear), and the abundances of light elements~\cite{Riess:1998cb,Perlmutter:1998np,Planck:2018vyg,ACT:2020gnv,eBOSS:2020yzd,KiDS:2020suj,Mossa:2020gjc,SPT-3G:2021eoc,DES:2021wwk,Scolnic:2021amr}.

Of course, this very economical picture does not need to be the end of the story as far as DE is concerned. One reason for this belief is the (in)famous \textit{cosmological constant problem}~\cite{Weinberg:1988cp}. In an attempt to construct more realistic physical models for DE, significant effort has been placed into going beyond the minimal scenario. Examples include models endowing DE with a dynamical nature~\cite{Wetterich:1987fm,Ratra:1987rm,Caldwell:1997ii}, or featuring non-trivial interactions between DE and other components of the Universe's energy budget. Examples of the latter are interacting DE (IDE) models, which exhibit interactions between dark matter (DM) and DE, historically motivated by attempts to address the \textit{coincidence problem}~\cite{Zlatev:1998tr,Huey:2004qv,Velten:2014nra}.\footnote{However, the amount of energy exchange required to address this problem is now excluded by observations. See in addition Refs.~\cite{DAmico:2016jbm,Marsh:2016ynw} for further issues which have been raised in this context owing to quantum corrections.}

State-of-the-art constraints on IDE cosmologies arise primarily from measurements of temperature and polarization anisotropies in the CMB, in combination with measurements of the late-time background expansion history from BAO and Hubble flow SNeIa. In combination, these probes have set stringent constraints on the strength of the DM-DE interaction, typically denoted by $\xi$: the constraints depend on the specific assumptions concerning the form of the interaction, and on the datasets adopted, but typically restrict $\vert \xi \vert \lesssim {\cal O}(0.1)$ (see e.g. Refs.~\cite{DiValentino:2017iww,Yang:2019uzo,Cheng:2019bkh,Yang:2020tax,Yang:2021oxc} for some of the latest constraints, see also Refs.~\cite{Zhang:2018mlj,Zhang:2018glx,Liu:2022hpz}). Incidentally, these tight constraints from BAO and SNeIa are the key reason why IDE scenarios fall short of fully addressing the Hubble tension~\cite{Bernal:2016gxb,Addison:2017fdm,Mortsell:2018mfj,Lemos:2018smw,Aylor:2018drw,Knox:2019rjx,Arendse:2019hev,Zhang:2020uan,Efstathiou:2021ocp,Cai:2021weh}, despite these models having experienced somewhat of a revival in recent years in attempts to address cosmological tensions (see e.g.\ Refs.~\cite{DiValentino:2020zio,DiValentino:2021izs,Perivolaropoulos:2021jda,Schoneberg:2021qvd,Abdalla:2022yfr} for reviews on the Hubble tension).

In recent years, significant efforts have gone into extracting LSS clustering information beyond that contained within the (reconstructed) BAO peaks, using information from the full-shape (FS) power spectrum of biased tracers of the LSS. These efforts have been driven in part by significant advances in the so-called Effective Field Theory of LSS (EFTofLSS)~\cite{Baumann:2010tm}, which have allowed for concrete applications of this framework to real data from galaxy redshift surveys: these advances motivate us to re-analyze IDE models in this context. It is our major goal in the present work to explore whether redshift-space galaxy clustering data can improve state-of-the-art constraints on the DM-DE interaction strength.\footnote{We note that the same interacting DE model that will be examined in this work had actually been confronted against FS galaxy power spectrum measurements by one of us in 2009 in Ref.~\cite{Gavela:2009cy}. However, this early study was performed using the FS power spectrum of luminous red galaxies from the Sloan Digital Sky Survey (SDSS) Data Release 4 (DR4) sample, alongside CMB and SNeIa measurements from WMAP and the Union dataset. All these datasets are now significantly outdated, and the modeling of the FS galaxy power spectrum was significantly simplified. Therefore it is extremely timely to re-examine this issue nearly 15 years later, making use of state-of-the-art cosmological measurements, in combination with a more robust theoretical modeling of the FS galaxy power spectrum. See also the later Ref.~\cite{Duniya:2015nva} for an investigation of horizon-scale relativistic effects in the galaxy power spectrum in the presence of interacting DE.}

The rest of this paper is then organized as follows. In Sec.~\ref{sec:theory}, we review the physics of IDE models, present the specific model we will study, before briefly discussing the EFTofLSS and how our model fits within this framework. In Sec.~\ref{sec:data} we discuss the datasets and analysis methodology we make use of. The results of this analysis are presented in Sec.~\ref{sec:results}. Finally, in Sec.~\ref{sec:conclusions} we provide our concluding remarks.

\section{Dark sector interactions and the full-shape galaxy power spectrum}
\label{sec:theory}

\subsection{Interacting dark energy}
\label{subsec:ide}

We begin by reviewing the basic features of IDE models. In the following,  we shall work under the assumption of a spatially flat Friedmann-Lema\^{i}tre-Robertson-Walker metric. In the absence of DM-DE interactions the stress-energy tensors of DM and DE, which we denote by $T^{\mu\nu}_c$ and $T^{\mu\nu}_x$ respectively, are separately covariantly conserved. This implies:
\begin{eqnarray}
\nabla_{\nu} T^{\mu\nu}_c = \nabla_{\nu} T^{\mu\nu}_x = 0\,,
\label{eq:tmunu0}
\end{eqnarray}
where $\nabla_{\mu}$ denotes the covariant derivative. However, particularly in the case where DE is described by a new light field (as in quintessence models), couplings of DE to matter fields (typically with gravitational strength) are somewhat unavoidable, unless protected by a specific symmetry~\cite{Carroll:1998zi}. When introducing non-gravitational interactions between DM and DE, there are essentially two ways to proceed. One can either work from first principles in the context of a specific (possibly UV-complete) model, where the interaction between the dark components is introduced at the level of the action. Historically, this was in fact the approach first followed, particularly in the context of so-called \textit{coupled quintessence} models, where a light quintessence DE field is coupled to the DM field~\cite{Wetterich:1994bg,Amendola:1999dr,Amendola:1999er,Mangano:2002gg,Farrar:2003uw}. Alternatively, one can introduce a phenomenological parametrization for the DM-DE interaction at the level of the conservation equations in Eq.~(\ref{eq:tmunu0}), in such a way that the two stress-energy tensors are separately not conserved, but their sum is. In this work, we shall follow the second approach.\footnote{We note that a separate but nonetheless associated possibility recently considered in the literature is that where DE interacts either with baryons~\cite{Vagnozzi:2019kvw,Jimenez:2020ysu,Vagnozzi:2021quy,Benisty:2021cmq,Ferlito:2022mok} or with electromagnetism~\cite{Calabrese:2013lga,Martins:2015ama,Martins:2015jta,Martins:2016oyv,Euclid:2021cfn}. See also Refs.~\cite{He:2017alg,Zhang:2021ygh} for works discussing the related possibility of direct detection of DE.} Typically, one assumes that the covariant derivatives of the DM and DE stress-energy tensors evolve as:
\begin{eqnarray}
\label{eq:conservDM}
\nabla_{\nu} T^{\mu\nu}_c  &=& \frac{Qu^{\mu}_c}{a}\,;\\
\label{eq:conservDE}
\nabla_{\nu} T^{\mu\nu}_x  &=& -\frac{Qu^{\mu}_c}{a}\,,
\end{eqnarray}
where $u^{\mu}_c$ is the DM velocity 4-vector, and $Q$ is the DM-DE interaction rate (with units of energy per volume per time). At this point, one needs to make a (phenomenological) choice for the functional form of $Q$. A choice commonly adopted in the literature is the following:
\begin{eqnarray}
Q = \xi{\cal H}\rho_x\,,
\label{eq:Q}
\end{eqnarray}
where ${\cal H}$ is the conformal Hubble rate, $\rho_x$ is the DE energy density and $\xi$ is a dimensionless parameter which controls the strength of the DM-DE interaction: a value of $\xi>0$ ($\xi<0$) indicates energy transfer from the DE (DM) to the DM (DE) sector.

A comment is in order regarding the appearance of ${\cal H}$ in the interaction rate. This might appear puzzling at first, as it may suggest that an interaction rate which should ultimately be determined by local interactions is actually sensitive to a global quantity such as the expansion rate, see e.g.\ Ref.~\cite{Valiviita:2008iv}. In reality, the appearance of ${\cal H}$ is ultimately a consequence of the first principle of thermodynamics, a local law (independently of the context to which one applies it, for example cosmology) which essentially states that changes in the density respond to changes in the volume due to cosmic expansion. In other words, the conservation equations do not explicitly know about the underlying cosmology or theory of gravity, but only about the change in scale or volume. In fact, one can eliminate ${\cal H}$ entirely and use the scale factor as time variable (as in the non-interacting case).\footnote{We thank the anonymous referee of one of our previous papers for drawing our attention to this issue, whereas S.V. thanks Marco Bruni for sharing this illuminating explanation.} In addition, we note that Ref.~\cite{Pan:2020zza} explicitly showed how interaction rates featuring factors related to ${\cal H}$, including but not limited to the case considered in Eq.~(\ref{eq:Q}), may naturally emerge from first principles when considering well-motivated field theories for IDE scenarios.

In the presence of the coupling given by Eq.~(\ref{eq:Q}), the continuity equations for the DM and DE energy densities $\rho_c$ and $\rho_x$ are modified to:
\begin{eqnarray}
\label{eq:continuitydm}
\dot{\rho}_c+3{\cal H}\rho_c &=& \xi{\cal H}\rho_x\,;\\
\label{eq:continuityde}
\dot{\rho}_x+3{\cal H}(1+w_x)\rho_x &=& -\xi{\cal H}\rho_x\,,
\label{eq:continuity}
\end{eqnarray}
where $w_x$ is the DE equation of state (EoS). Assuming that both $w_x$ ad $\xi$ are constant in cosmic time, the above Eqs.~(\ref{eq:continuitydm},\ref{eq:continuityde}) can be analytically integrated to give:
\begin{eqnarray}
\label{eq:solutionrhoc}
\rho_c &=& \frac{\rho_{c,0}}{a^3}+\frac{\rho_{x,0}}{a^3} \left [ \frac{\xi}{3w_x+\xi} \left ( 1-a^{-3w_x-\xi} \right ) \right ]\,; \\
\label{eq:solutionrhox}
\rho_x &=& \frac{\rho_{x,0}}{a^{3(1+w_x)+\xi}} \,,
\end{eqnarray}
where $\rho_{c,0}$ and $\rho_{x,0}$ are the present-day energy densities of DM and DE. From Eq.~(\ref{eq:solutionrhox}), we see that in the presence of such an interaction the DE component effectively behaves as a component with EoS $w_{x,{\rm eff}}=w_x+\xi/3$.

The presence of the DM-DE coupling also modifies the evolution of perturbations. Working in synchronous gauge, the coupled system of linear Einstein-Boltzmann equations for the evolution of the DM and DE density perturbations ($\delta_c$ and $\delta_x$) and velocity divergences ($\theta_c$ and $\theta_x$) is given by~\cite{Valiviita:2008iv,Gavela:2010tm,LopezHonorez:2010esq}:
\small
\begin{eqnarray}
\label{eq:deltac}
\dot{\delta}_c &=& -\theta_c - \frac{1}{2}\dot{h} +\xi{\cal H}\frac{\rho_x}{\rho_c}(\delta_x-\delta_c)+\xi\frac{\rho_x}{\rho_c} \left ( \frac{kv_T}
{3}+\frac{\dot{h}}{6} \right )\,; \\
\label{eq:thetac}
\dot{\theta}_c &=& -{\cal H}\theta_c\,;\\
\label{eq:deltax}
\dot{\delta}_x &=& -(1+w_x) \left ( \theta_x+\frac{\dot{h}}{2} \right )-\xi \left ( \frac{kv_T}{3}+\frac{\dot{h}}{6} \right ) \nonumber \\
&&-3{\cal H}(1-w_x) \left [ \delta_x+\frac{{\cal H}\theta_x}{k^2} \left ( 3+3w_x+\xi \right ) \right ]\,;\\
\label{eq:thetax}
\dot{\theta}_x &=& 2{\cal H}\theta_x+\frac{k^2}{1+w_x}\delta_x+2{\cal H}\frac{\xi}{1+w_x}\theta_x-\xi{\cal H}\frac{\theta_c}{1+w_x}\,,
\end{eqnarray}
\normalsize
where we set the DE sound speed squared to $c_{s,x}^2=1$, and $h$ and $v_T$ refer to the trace of the metric perturbation $h_{ij}$ in the synchronous gauge and to the center of mass velocity for the total fluid respectively, where the appearance of the latter is required by gauge invariance arguments~\cite{Gavela:2010tm}. The initial conditions for the DE density perturbation and velocity divergence are also modified, following Ref.~\cite{Gavela:2010tm}.

Furthermore, one needs to avoid instabilities in the system of Eqs.~(\ref{eq:deltac}--\ref{eq:thetax}). Gravitational and non-adiabatic (early-time) instabilities can be avoided provided that \textit{a)} $w_x \neq -1$, and \textit{b)} $\xi$ and $(1+w_x)$ carry opposite signs~\cite{Gavela:2009cy,Gavela:2010tm}. In this work, our goal will be that of examining a scenario which is as close as possible to that of an interacting vacuum, where $w_x=-1$. This is strictly speaking not possible, due to gravitational instabilities. However, here we will opt for the phenomenological choice of setting $1+w_x$ to a small but non-zero value. Specifically, we fix $w_x=-0.999$, thus requiring $\xi<0$. The rationale behind this approach is that, for $1+w_x$ sufficiently small, the effect of the DE perturbations is negligible at the level of the coupled Einstein-Boltzmann system, which are instead essentially only capturing the effect of the DM-DE interaction governed by $\xi$. This was explicitly demonstrated by one of us through simulated data in Ref.~\cite{DiValentino:2020leo}.\footnote{In principle, we could equally well have opted for the choice of setting $w_x=-1.001$ and $\xi>0$. However, there are at least two good reasons to not consider this possibility. The first is that quintessence-like DE scenarios where $w_x>-1$ are actually theoretically more motivated (or at least theoretically easier to come about) compared to phantom scenarios where $w_x<-1$. The second is that a scenario where energy flows from the DM to the DE ($\xi<0$) is somewhat theoretically more natural than the reverse case ($\xi>0$). Note, however, that the simplest quintessence models appear on their own to be observationally disfavored, as they worsen the $H_0$ tension~\cite{Banerjee:2020xcn}.} We note that such an approach has already been followed in several previous works, see e.g. Refs.~\cite{DiValentino:2019ffd,DiValentino:2019jae,Lucca:2020zjb}. We also note that another approach towards avoiding gravitational instabilities, investigated in Refs.~\cite{Li:2014eha,Li:2014cee,Guo:2017hea,Zhang:2017ize,Guo:2018gyo}, is to extend the parameterized post-Friedmann approach to the IDE case.

In closing, it is worth noting that IDE models have received significant attention in recent years. For a selection of important works on various aspects of IDE models besides those already discussed, ranging from model-building to structure formation simulations to observational constraints, we refer the reader to Refs.~\cite{Amendola:2003eq,Pettorino:2004zt,Pettorino:2005pv,Barrow:2006hia,Bean:2007ny,He:2008tn,Pettorino:2008ez,Baldi:2008ay,Majerotto:2009np,Jamil:2009eb,Martinelli:2010rt,Baldi:2010td,DeBernardis:2011iw,Baldi:2011th,Pettorino:2012ts,Carbone:2013dna,Piloyan:2013mla,Pettorino:2013oxa,Pourtsidou:2013nha,Piloyan:2014gta,Faraoni:2014vra,Ferreira:2014cla,Planck:2015bue,Skordis:2015yra,Tamanini:2015iia,Marra:2015iwa,Murgia:2016ccp,Pourtsidou:2016ico,Nunes:2016dlj,Kumar:2016zpg,Kumar:2017dnp,Benisty:2017eqh,Yang:2017zjs,Mifsud:2017fsy,Yang:2017ccc,Kumar:2017bpv,Guo:2017deu,Dutta:2017fjw,Feng:2017usu,Benisty:2018qed,Barros:2018efl,Costa:2018aoy,Yang:2018euj,vonMarttens:2018iav,Elizalde:2018ahd,Guo:2018ans,Yang:2018uae,Li:2018ydj,vonMarttens:2018bvz,Cardenas:2018nem,Benisty:2018oyy,Bonici:2018qli,Martinelli:2019dau,Kumar:2019wfs,Pan:2019jqh,Li:2019loh,Yang:2019vni,Pan:2019gop,Landim:2019lvl,Benetti:2019lxu,Carneiro:2019rly,Kase:2019veo,Liu:2019ygl,Yang:2019uog,Chamings:2019kcl,Sharma:2020glf,Yang:2020uga,Hogg:2020rdp,Amendola:2020ldb,Benisty:2020nql,Gomez-Valent:2020mqn,Aljaf:2020eqh,DiValentino:2020evt,Mukhopadhyay:2020bml,DiValentino:2020vnx,DiValentino:2020kpf,Yao:2020pji,BeltranJimenez:2020qdu,Yang:2021hxg,Sinha:2021tnr,Gao:2021xnk,Zhang:2021yof,Salzano:2021zxk,Wang:2021kxc,Benetti:2021div,Kumar:2021eev,Figueruelo:2021elm,Bora:2021uxq,Jimenez:2021ybe,Carrilho:2021hly,Lucca:2021eqy,Bora:2021iww,Linton:2021cgd,Nunes:2021zzi,Anchordoqui:2021gji,Hogg:2021yiz,Guo:2021rrz,Alestas:2021luu,Mancini:2021lec,Gariazzo:2021qtg,Carrilho:2021rqo,Sharma:2021ivo,Duniya:2022miz} and to the review of Ref.~\cite{Wang:2016lxa}.

\subsection{Interacting dark energy and the Effective Field Theory of Large-Scale Structure}
\label{subsec:ideeftoflss}

There is a wealth of information contained in the linear and mildly non-linear clustering of tracers of the LSS, such as galaxies, on which our analysis will focus on. In recent years, significant effort has been devoted to extract information beyond that contained within the (reconstructed) BAO peaks~\cite{Eisenstein:2006nk,Sherwin:2012nh}, considering the Full-Shape (broadband) redshift-space galaxy power spectrum. Nevertheless, in order to fully exploit current and future FS redshift-space galaxy power spectra measurements, a robust theoretical modeling is mandatory~\cite{Bose:2017jjx,FonsecadelaBella:2018leo,Osato:2018ldv,Bose:2019psj,Bose:2019ywu}. Various approaches towards such a robust modeling exist in the literature. Here, we shall make use of the EFTofLSS approach~\cite{Baumann:2010tm}, see Ref.~\cite{Cabass:2022avo} for a recent comprehensive review. While, in the very beginning, one of the most popular theoretical modeling approaches was Standard Perturbation Theory (SPT)~\cite{Scoccimarro:1995if}, the EFTofLSS can be regarded as the final product of a consistent perturbation theory approach to describe the mildly non-linear clustering of tracers of the LSS: for an incomplete list of other approaches developed throughout this evolution, see e.g. Refs.~\cite{Zeldovich:1969sb,Crocce:2005xy,Bernardeau:2008fa,Matsubara:2008wx,Taruya:2010mx,Bernardeau:2011dp,Carlson:2012bu,Vlah:2016bcl,Hand:2017ilm,Chen:2020zjt}.

The EFTofLSS is a LSS perturbation theory approach (where one can think of the expansion variable as being the overdensity field smoothed over an appropriate scale) which can be used to robustly model the mildly non-linear clustering of LSS tracers~\cite{Baumann:2010tm}. In a nutshell, the EFTofLSS provides a framework to  characterize the back-reaction and impact of unknown or poorly-known short-scale physics, such as the complex details of galaxy formation, on long-wavelength modes: this is achieved through a set of additional counter-terms, whose functional form is fully fixed once the symmetries obeyed by the LSS field are specified, and whose amplitude is controlled by a set of free coefficients of unknown magnitude which, lacking a precise knowledge of the short-wavelength physics in question, have to be treated as nuisance parameters to be fit to the data. When combined with an IR resummation procedure to treat long-wavelength displacements (i.e.\ bulk motions, whose effect is crucial to properly model the non-linear evolution of the BAO peak), stochastic contributions, non-linear biasing terms, and Alcock-Paczynski effects, one obtains the most general, symmetry-driven model for the mildly non-linear clustering of biased LSS tracers such as galaxies, which integrates out the complex and poorly-known details of short-scale (UV) physics.

Over the past decade, significant work has gone into developing the EFTofLSS from its initial formulation, to the point of enabling practical applications to real data from current galaxy surveys. While several of the intermediate results are not, strictly speaking, required to analyze galaxy clustering data, much as the ``West Coast'' EFTofLSS team we believe that it is only fair to acknowledge key developments on the EFTofLSS theory~\cite{McDonald:2009dh,Carrasco:2012cv,Pajer:2013jj,Carrasco:2013sva,Mercolli:2013bsa,Carrasco:2013mua,Carroll:2013oxa,Porto:2013qua,Senatore:2014via,Baldauf:2014qfa,Angulo:2014tfa,Senatore:2014eva,Senatore:2014vja,Lewandowski:2014rca,Mirbabayi:2014zca,McQuinn:2015tva,Foreman:2015uva,Angulo:2015eqa,Baldauf:2015xfa,Assassi:2015jqa,Baldauf:2015tla,Baldauf:2015zga,Baldauf:2015aha,Foreman:2015lca,Abolhasani:2015mra,Assassi:2015fma,Lewandowski:2015ziq,Bertolini:2015fya,Bertolini:2016bmt,Blas:2016sfa,Cataneo:2016suz,Bertolini:2016hxg,Fujita:2016dne,Perko:2016puo,Lewandowski:2016yce,Lewandowski:2017kes,Senatore:2017hyk,Simonovic:2017mhp,Senatore:2017pbn,Nadler:2017qto,Bose:2018orj,deBelsunce:2018xtd,Lewandowski:2018ywf,Konstandin:2019bay,Nishimichi:2020tvu,Donath:2020abv,Steele:2020tak,Braganca:2020nhv,Steele:2021lnz,Baldauf:2021zlt} and data~\cite{DAmico:2019fhj,Ivanov:2019pdj,Colas:2019ret,Ivanov:2019hqk,Philcox:2020vvt,DAmico:2020kxu,Chudaykin:2020aoj,Ivanov:2020ril,DAmico:2020ods,Wadekar:2020hax,Chudaykin:2020hbf,Chudaykin:2020ghx,DAmico:2020tty,Ivanov:2021zmi,Ivanov:2021fbu,DAmico:2021ymi,Ivanov:2021kcd,Philcox:2021kcw,Cabass:2022wjy,DAmico:2022gki} sides in any work which applies the EFTofLSS to real data.\footnote{For a more complete accounting of the theoretical developments associated to Refs.~\cite{McDonald:2009dh,Carrasco:2012cv,Pajer:2013jj,Carrasco:2013sva,Mercolli:2013bsa,Carrasco:2013mua,Carroll:2013oxa,Porto:2013qua,Senatore:2014via,Baldauf:2014qfa,Angulo:2014tfa,Senatore:2014eva,Senatore:2014vja,Lewandowski:2014rca,Mirbabayi:2014zca,McQuinn:2015tva,Foreman:2015uva,Angulo:2015eqa,Baldauf:2015xfa,Assassi:2015jqa,Baldauf:2015tla,Baldauf:2015zga,Baldauf:2015aha,Foreman:2015lca,Abolhasani:2015mra,Assassi:2015fma,Lewandowski:2015ziq,Bertolini:2015fya,Bertolini:2016bmt,Blas:2016sfa,Cataneo:2016suz,Bertolini:2016hxg,Fujita:2016dne,Perko:2016puo,Lewandowski:2016yce,Lewandowski:2017kes,Senatore:2017hyk,Simonovic:2017mhp,Senatore:2017pbn,Nadler:2017qto,Bose:2018orj,deBelsunce:2018xtd,Lewandowski:2018ywf,Konstandin:2019bay,Nishimichi:2020tvu,Donath:2020abv,Steele:2020tak,Braganca:2020nhv,Steele:2021lnz,Baldauf:2021zlt}, we refer the reader to what by now is a standard footnote in all the papers by the ``West Coast'' EFTofLSS team, for instance footnote~1 in Ref.~\cite{DAmico:2022gki}.}

In this work, we shall adopt the EFTofLSS to model the mildly non-linear full-shape galaxy power spectrum in the presence of the interactions between DM and DE previously described in Sec.~\ref{subsec:ide}. As the complete set of equations for the redshift-space galaxy power spectrum is rather cumbersome, we will not report it here, but refer the reader to Ref.~\cite{Chudaykin:2020aoj}, where the model and its implementation in the \texttt{CLASS-PT} Boltzmann solver are described in detail. In our case, the only ingredient of the underlying EFTofLSS model we need to modify is the linear power spectrum $P_{\rm lin}(k,z)$, which we compute as usual by evolving the linear Einstein-Boltzmann equations given in Eqs.~(\ref{eq:deltac}--\ref{eq:thetax}). In turn, $P_{\rm lin}(k,z)$ enters in the convolution integrals required to compute the 1-loop SPT power spectrum $P_{\rm 1-loop,SPT}(k,z)$, and is also required to compute the counter-term $P_{\rm ctr}(k,z)$, needed for consistency of the 1-loop result, whose amplitude is modulated by the effective sound speed parameter. We keep the functional form of the SPT kernels, counter-terms, bias expansion, redshift-space kernels, and stochastic contributions unchanged, while the implementation of IR resummation is also left unchanged.

A comment is in order regarding the fact that we only need to modify the input linear power spectrum, which is the only place where new physics due to DM-DE interactions enters. While the default implementation of \texttt{CLASS-PT} is targeted towards the $\Lambda$CDM model, one can essentially envisage extending the underlying model to three classes of new fundamental physics scenarios with increasing levels of complexity:
\begin{enumerate}
\item New physics which only affects the expansion, thermal history, and linear evolution of perturbations. In this case, no changes to the standard routines are required, as these scenarios only modify the ingredients which are already present, such as the linear power spectrum, growth factor, and so on.
\item New physics which modifies the mode-coupling kernels while preserving the equivalence principle. Several modified gravity models fall within this class, in which case one needs to recompute the perturbation theory matrices incorporating the new kernels, required to compute the 1-loop power spectrum. The specific details will of course depend on the modified gravity model, see e.g. Ref.~\cite{Cusin:2017wjg}.
\item New physics which modifies the mode-coupling kernels \textit{and} violates the equivalence principle, as also occurring in certain modified gravity models. In this case, one needs to alter the IR resummation procedure accordingly~\cite{Creminelli:2013poa,Crisostomi:2019vhj,Lewandowski:2019txi}.
\end{enumerate}
As we consider a purely phenomenological implementation of the DM-DE interactions at the level of linear Einstein-Boltzmann equations, without committing to any specific UV-complete underlying field theory model which describes such interactions, the new physics explored here belongs to the first level described above. Given the size of the uncertainties associated with current galaxy clustering measurements, our approach can be viewed as a conservative one. In closing, we also note that the same approach has been adopted by several other works in the recent literature, where both early- and late-time new physics scenarios have been constrained against FS galaxy clustering measurements using the EFTofLSS implementation of either \texttt{CLASS-PT}~\cite{Chudaykin:2020aoj} or \texttt{PyBird}~\cite{DAmico:2020kxu}, with the only modifications being the input linear matter power spectra (see e.g.\ Refs.~\cite{Niedermann:2020qbw,Chudaykin:2020igl,Braglia:2020auw,Allali:2021azp,Lague:2021frh,Jiang:2021bab,Ye:2021iwa,Xu:2021rwg,Herold:2021ksg,Chudaykin:2022rnl}).

\begin{figure*}
\begin{center}
\includegraphics[width=8cm]{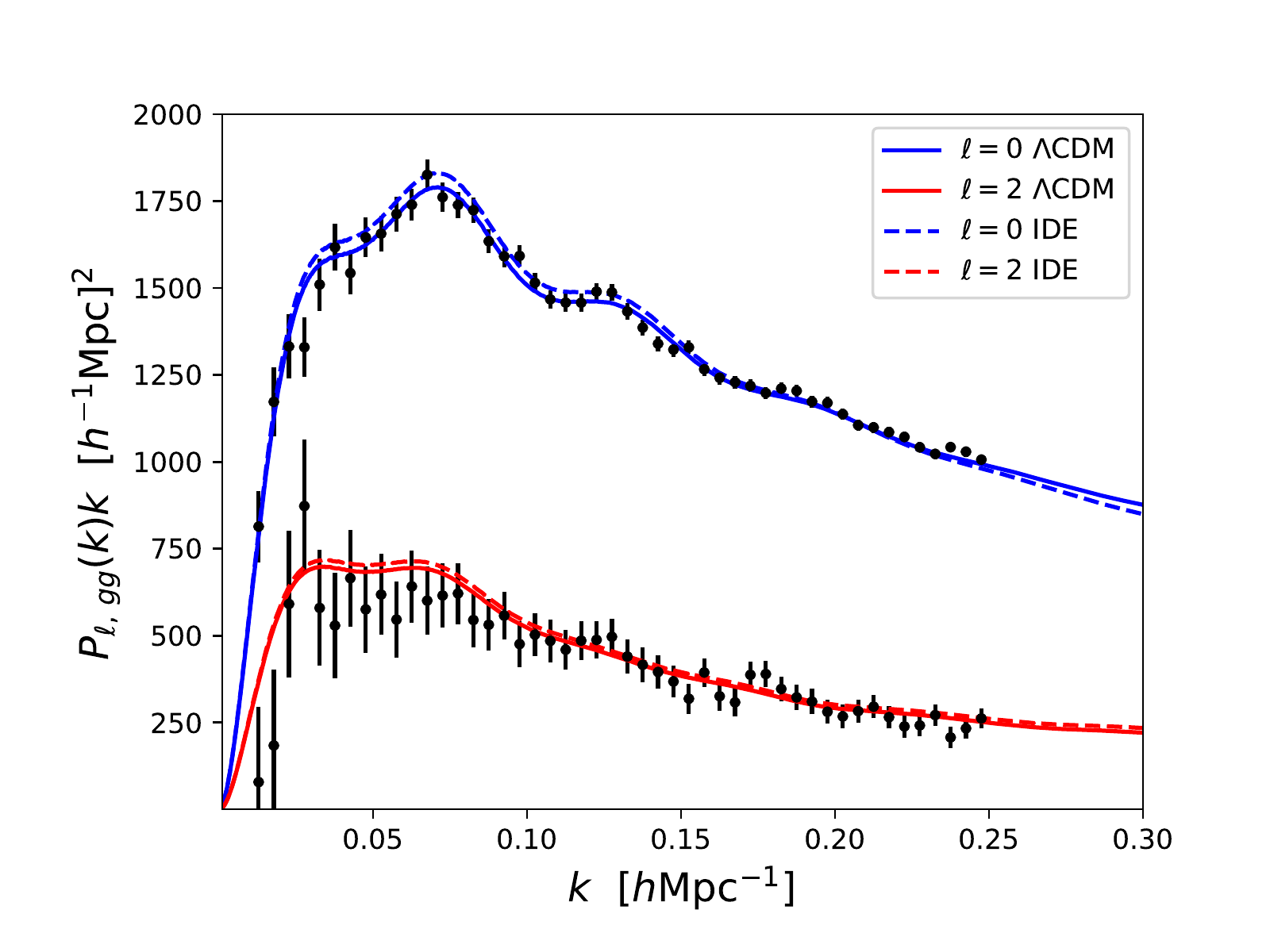} 
\includegraphics[width=8cm]{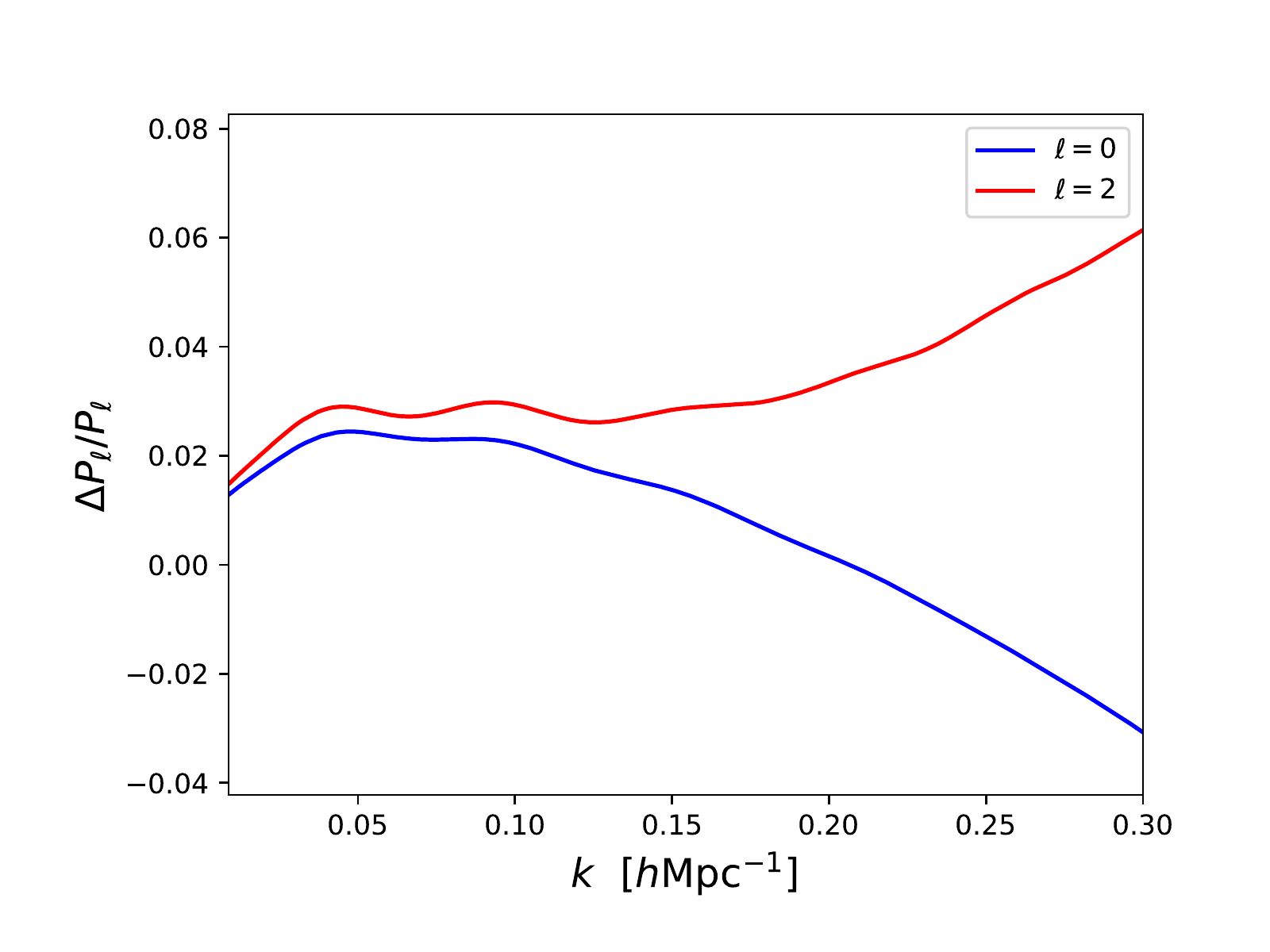} 
\caption{\textit{Left panel}: measurements of the monopole ($\ell=0$) and quadrupole ($\ell=2$) of the full-shape power spectrum of the BOSS DR12 North galactic cap galaxy field at the effective redshift $z_{\rm eff}=0.61$ correspond to the black datapoints. The blue and red curves correspond to the theoretical predictions for the monopole and quadrupole respectively, for the $\Lambda$CDM (solid) and IDE (dashed) models. Cosmological parameters are fixed to their best-fit values obtained from the \textit{CMB}+\textit{BAO}+\textit{FS} dataset combination (see Sec.~\ref{sec:data} for a discussion of these datasets), with the DM-DE coupling strength $\xi$ instead being fixed to $\xi=-0.05$, whereas nuisance parameters are fixed to their $\Lambda$CDM best-fit values. Note that for both the monopole and quadrupole what is being plotted is not $P_{\ell}(k)$, but $kP_{\ell}(k)$. \textit{Right panel}: relative difference $\Delta P_{\ell}/P_{\ell}$ between the IDE and $\Lambda$CDM monopole (blue) and quadrupole (red), with the same choice of cosmological and nuisance parameters as in the left panel.}
\label{fig:residuals}
\end{center}
\end{figure*}

The effects of IDE on the linear matter power spectrum have been amply discussed in Ref.~\cite{Lucca:2021dxo} (to which we refer the reader for more in-depth discussions), which however did not carry out the comparison against FS measurements we will instead perform. Introducing DM-DE interactions leads to two important effects. Notice from Eq.~(\ref{eq:solutionrhoc}) that in the presence of DM-DE interactions with strength $\xi<0$ (for a fixed value of $\rho_{c,0}$) the amount of DM grows with respect to its $\Lambda$CDM counterpart when going back in time. This on its own would result in a higher redshift of matter-radiation equality, and therefore a larger equality wavenumber since $k_{\rm eq} \propto \sqrt{\omega_m(1+z_{\rm eq})}$. However, as we shall see later, IDE introduces correlations between parameters which result in parameter shifts once $\xi \neq 0$, most notably a decrease in both $\omega_c$ and $\Omega_m$. Accounting for these parameter shifts from a fit to the full CMB+BAO+FS dataset as we will discuss later, we see that the overall effect is to slightly decrease $z_{\rm eq}$, from the $\Lambda$CDM value $z_{\rm eq} \approx 3387$ to $z_{\rm eq} \approx 3330$ (the shift is larger if we were to consider the best-fit CMB+FS or CMB+BAO values, changing to $z_{\rm eq} \approx 2984$ and $z_{\rm eq} \approx 2819$ respectively).

Recall that the equality wavenumber $k_{\rm eq}$ determines the turnaround scale in the matter power spectrum (as the growth of modes with $k>k_{\rm eq}$, which entered the horizon during radiation domination, was suppressed by the large radiation pressure, which instead did not hinder the growth of modes with $k<k_{\rm eq}$). Note that the shifts in $z_{\rm eq}$ and therefore $k_{\rm eq}$ discussed previously result in slight changes in the horizon scale. In addition, for $\xi<0$, the conformal time today $\eta_0$ increases, leading to a small overall increase in the large-scale (small-$k$) amplitude of the power spectrum. Another feature will be that due the onset of DE domination which, for $\xi<0$, occurs earlier, resulting in a small suppression of the power spectrum on smaller scales (which however are beyond the scales we will probe in our analysis).

The theoretical predictions for the monopole ($\ell=0$) and quadrupole ($\ell=2$) of the full-shape power spectrum within $\Lambda$CDM and the IDE model considered here are shown in Fig.~\ref{fig:residuals}. We have fixed the cosmological parameters to the best-fit values obtained from a combined analysis of CMB, pre-reconstructed FS, and post-reconstruction BAO data (see details in Secs.~\ref{sec:data} and \ref{sec:results}), whereas we have fixed the DM-DE coupling strength to $\xi=-0.05$. Note that we have kept nuisance parameters fixed to their $\Lambda$CDM best-fit values when moving across the two models, in order to focus exclusively on the effects induced by the cosmological parameters. In reality, nuisance parameters shift slightly when moving from one model to another, and as a result absorb part of the differences shown in Fig.~\ref{fig:residuals}. However, we have explicitly checked that the shifts in all nuisance parameters remain below the $1\sigma$ level, and that the extent to which they will absorb part of the signal of interest is rather limited. Notice from Fig.~\ref{fig:residuals} that the cleanest signature of IDE appears to be in the monopole. Already from these qualitative considerations we can expect that broadband information in the FS galaxy power spectrum, and in particular information related to the equality scale, will play an important role in setting constraints on IDE. Our analysis will confirm these expectations, as we will later discuss in Sec.~\ref{sec:results}.

\section{Datasets and methodology}
\label{sec:data}

In order to set constraints on the parameters of the IDE model we shall consider various combinations of the following datasets:
\begin{itemize}
\item Measurements of CMB temperature anisotropy and polarization power spectra, as well as their cross-spectra, from the \textit{Planck} 2018 legacy data release. We consider the high-$\ell$ \texttt{Plik} likelihood for TT (in the multipole range $30 \leq \ell \leq 2508$) as well as TE and EE (in the multipole range $30 \leq \ell \leq 1996$), in combination with the low-$\ell$ TT-only ($2 \leq \ell \leq 29$) likelihood based on the \texttt{Commander} component-separation algorithm in pixel space, as well as the low-$\ell$ EE-only ($2 \leq \ell \leq 29$) \texttt{SimAll} likelihood~\cite{Planck:2019nip}. We also include measurements of the CMB lensing power spectrum, as reconstructed from the temperature 4-point function~\cite{Planck:2018lbu}. We refer to this dataset combination as \textit{\textbf{CMB}}.
\item Post-reconstruction Baryon Acoustic Oscillation measurements from the BOSS DR12~\cite{BOSS:2016wmc} survey. We refer to this dataset as \textit{\textbf{BAO}}.
\item Measurements of the monopole and quadrupole ($\ell=0$ and $\ell=2$ respectively) of the full-shape power spectrum of the BOSS DR12 galaxies, divided into four independent inputs: two distinct redshift bins at $z_{\rm eff}=0.38$ and $z_{\rm eff}=0.61$, observed in the North and South galactic caps (NGC and SGC respectively). We refer to this dataset combination as \textit{\textbf{FS}}.
\item A Gaussian prior on the physical baryon density parameter $\omega_b$ arising from Big Bang Nucleosynthesis (BBN) constraints on the abundance of light elements: $100\omega_b =  2.233 \pm 0.036$~\cite{Mossa:2020gjc}. We refer to this prior as \textit{\textbf{BBN}}.
\end{itemize}
For more details regarding window function, post-reconstruction BAO measurements extraction, BAO-FS covariance matrix, and implementation of the joint FS+BAO likelihood, we refer the reader to Refs.~\cite{Ivanov:2019pdj,Philcox:2020vvt,Chudaykin:2020aoj}, which discusses these issues in depth.

Model-wise, we consider a seven-parameter IDE model, where the strength of the DM-DE interaction $\xi$ is allowed to vary alongside the six standard $\Lambda$CDM parameters. We set a flat prior on $\xi \in [-1;0]$  in order to avoid gravitational and early-time non-adiabatic instabilities (see Sec.~\ref{subsec:ide}), verifying a posteriori that our results are not affected by the choice of lower prior boundary. We set the default, wide conservative priors on the EFTofLSS nuisance parameters following Ref.~\cite{Ivanov:2019pdj}. As we are including CMB measurements, we do not fix either the scalar spectral index $n_s$ nor the baryon-to-dark-matter density ratio $\omega_b/\omega_c$, given that both these parameters are extremely well constrained by CMB data.

Theoretical predictions for the relevant observables are obtained using the Boltzmann solver \texttt{CLASS-PT}~\cite{Chudaykin:2020aoj}, which is itself an extension of the Boltzmann solver \texttt{CLASS}~\cite{Blas:2011rf,Lesgourgues:2011re}, and allows to compute the 1-loop auto- and cross-power spectra for matter fields and biased tracers both in real and redshift space, incorporating all the ingredients discussed in Sec.~\ref{subsec:ideeftoflss} required for the comparison to data. Our theoretical model for the FS measurements is based on the EFTofLSS predictions at 1-loop order (corresponding to perturbative order $n=3$), and we consider FS measurements in the wavenumber range $k \in [0.01;0.2]\,h{\rm Mpc}^{-1}$ (slightly more conservative than the wavenumber range adopted in other works which have used the very same likelihood). As explained earlier, we only modify the input linear power spectrum obtained by solving the system of coupled linear Einstein-Boltzmann equations for the IDE model given in Eqs.~(\ref{eq:deltac}--\ref{eq:thetax}), without altering the SPT kernels, the structure of the counter-terms and stochastic contributions, the IR resummation implementation, and so on.\footnote{Note that, besides the EFTofLSS, a number of other theoretical modeling approaches have been adopted in analysis and forecasts for galaxy clustering data. Some of these studies considered simpler phenomenological and/or theory-motivated models for the FS power spectrum and higher order correlators, whereas others considered compressed versions thereof (see e.g.\ Refs.~\cite{Escudero:2015yka,Cuesta:2015iho,Giusarma:2016phn,BOSS:2016teh,BOSS:2016off,BOSS:2016hvq,BOSS:2016psr,Doux:2017tsv,Gualdi:2017iey,Sprenger:2018tdb,Gil-Marin:2018cgo,Giusarma:2018jei,Vagnozzi:2018pwo,Loureiro:2018qva,Gualdi:2019ybt,Troster:2019ean,Chen:2020fxs,Kobayashi:2020zsw,Gil-Marin:2020bct,Gualdi:2020eag,Cuceu:2021hlk,Gualdi:2021yvq,Brieden:2021edu,Brieden:2021cfg,Chen:2021wdi,Semenaite:2021aen,Neveux:2022tuk,Gualdi:2022kwz,Brieden:2022ieb,Gil-Marin:2022hnv}).}

We sample the posterior distributions for the parameters of the IDE model through Monte Carlo Markov Chain (MCMC) methods, using the cosmological sampler \texttt{MontePython}~\cite{Audren:2012wb,Brinckmann:2018cvx}. We assess the convergence of the MCMC chains using the Gelman-Rubin parameter $R-1$~\cite{Gelman:1992zz}, requiring $R-1<0.001$ for the chains to be considered converged.

When considering analyses involving the \textit{BAO} dataset but not the \textit{FS} one (e.g.\ when considering the \textit{CMB}+\textit{BAO} dataset combination), we make use of the ``standard'' non-EFTofLSS likelihood which is publicly available at \href{https://github.com/brinckmann/montepython\_public}{github.com/brinckmann/montepython\_public} (in particular, we use the \texttt{bao\_boss\_dr12} likelihood and not the \texttt{bao\_fs\_boss\_dr12} one, with the latter incorporating also $f\sigma_8$ measurements). When instead combining the \textit{BAO} and \textit{FS} datasets, we make use of the joint BAO-FS likelihood which is publicly available at \href{https://github.com/oliverphilcox/full\_shape\_likelihoods}{github.com/oliverphilcox/full\_shape\_likelihoods}, see Ref.~\cite{Philcox:2020vvt}. This strategy will provide the key to isolate the advantages of using the joint BAO-FS likelihood, and in particular the inclusion of FS measurements in the context of IDE models. In fact, thanks to our choice, our analyses involving the \textit{BAO} dataset but not the \textit{FS} one will be as close as possible to pre-2020 analyses which preceded the public release of the joint BAO-FS EFTofLSS likelihood we are using.\footnote{Recall that state-of-the-art constraints on IDE models mostly arise from the combination of CMB and geometrical information from the reconstructed BAO peaks, see for instance Ref.~\cite{DiValentino:2019jae}.}

\begin{table*}[!t]
\centering
\scalebox{1.2}{
\begin{tabular}{|c||ccc|}       
\hline\hline
Parameter & \textit{CMB}+\textit{BAO} & \textit{CMB}+\textit{FS} & \textit{CMB}+\textit{BAO}+\textit{FS} \\ \hline
$\omega_c$ & $0.094^{+0.022}_{-0.010}$ & $0.101^{+0.015}_{-0.009}$ & $0.115^{+0.005}_{-0.001}$ \\
$\xi$ & $-0.22^{+0.18}_{-0.09}\, [>-0.48]$ & $>-0.35$ & $>-0.12$ \\
\hline
$H_0\,[{\rm km}/{\rm s}/{\rm Mpc}]$ & $69.55^{+0.98}_{-1.60}$ & $69.04^{+0.84}_{-1.10}$ & $68.02^{+0.49}_{-0.60}$ \\
$\Omega_m$ & $0.243^{+0.054}_{-0.030}$ & $0.261^{+0.038}_{-0.025}$ & $0.299^{+0.015}_{-0.007}$ \\
\hline \hline     
\end{tabular}}
\caption{Constraints on selected parameters of the seven-parameter IDE model, obtained from various dataset combinations as indicated in the upper section of the Table, all involving the \textit{CMB} dataset. For the DM-DE coupling strength $\xi$ we report 95\%~C.L. lower limits, unless the 68\%~C.L. interval is consistent with a detection, in which case we report the latter, alongside the 95\%~C.L. in square brackets (here this situation only occurs when considering the \textit{CMB}+\textit{BAO} dataset combination), whereas for all other cosmological parameters we report 68\%~C.L. intervals. The final two rows are separated from the previous rows to highlight the fact that both $H_0$ and $\Omega_m$ are derived parameters.}
\label{tab:parametersidecmb}
\end{table*}

\begin{table*}[!t]
\centering
\scalebox{1.2}{
\begin{tabular}{|c||cc|}       
\hline\hline
Parameter & \textit{FS}+\textit{BAO} & \textit{FS}+\textit{BAO}+\textit{BBN} \\ \hline
$\omega_c$ & $0.100 \pm 0.023$ & $0.104^{+0.021}_{-0.017}$ \\
$\xi$ & $-0.31^{+0.20}_{-0.13}\,[>-0.62]$ & $>-0.62$ \\
\hline
$H_0\,[{\rm km}/{\rm s}/{\rm Mpc}]$ & $70.7^{+5.5}_{-4.9}$ & $71.8^{+1.6}_{-1.9}$ \\
$\Omega_m$ & $0.245^{+0.049}_{-0.036}$ & $0.257^{+0.044}_{-0.031}$ \\
\hline \hline                                                  
\end{tabular}}
\caption{As Tab.~\ref{tab:parametersidecmb}, but considering dataset combinations not involving the \textit{CMB} dataset. Note that we report the 68\%~C.L. interval on $\xi$ when considering the \textit{FS}+\textit{BAO} dataset combination, as it is consistent with a detection, with the 95\%~C.L. lower limit given in square brackets.}
\label{tab:parametersidenocmb}
\end{table*}

\section{Results}
\label{sec:results}

\begin{figure*}
\begin{center}
\includegraphics[width=15cm]{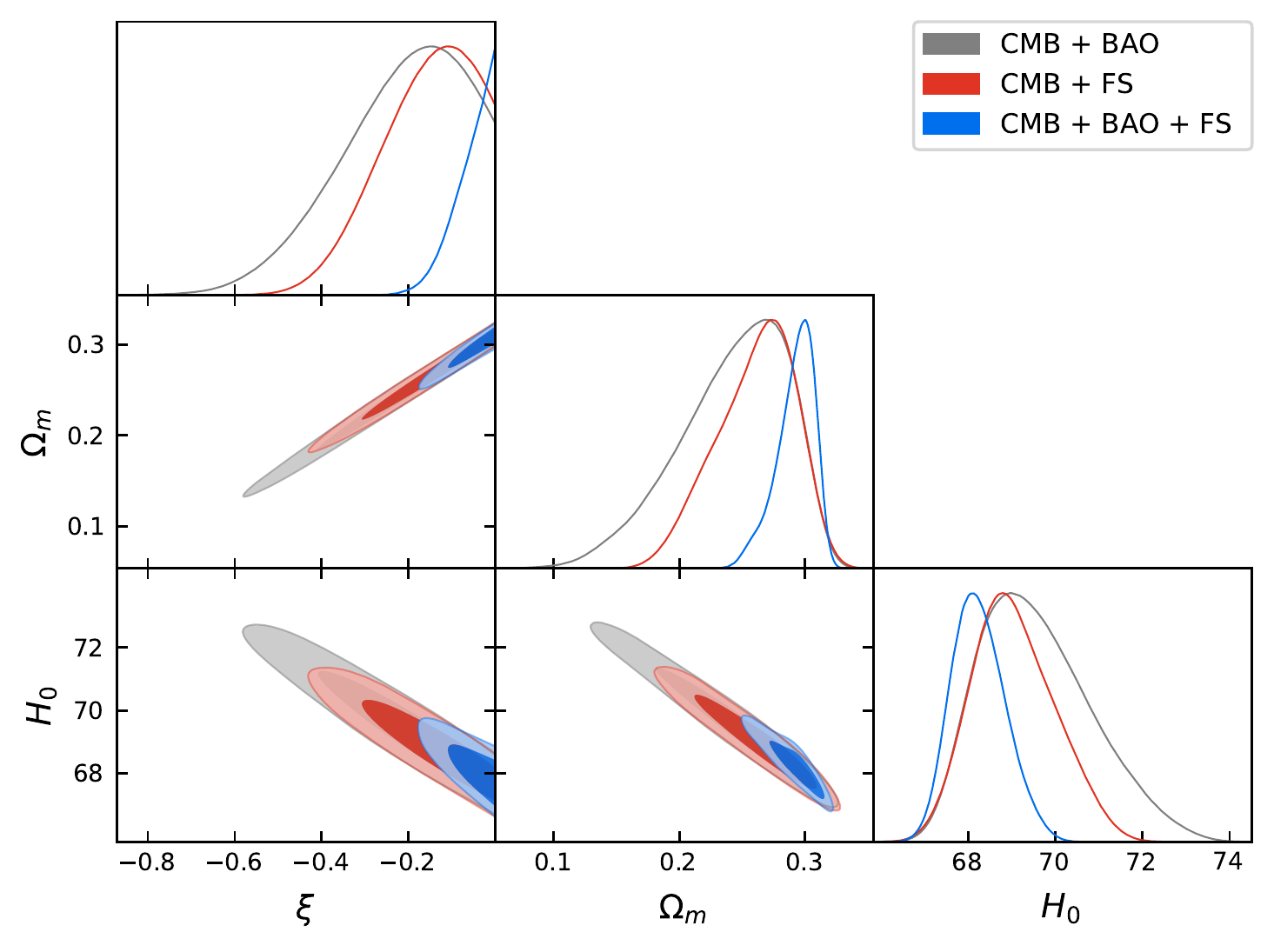} 
\caption{Triangular plot showing the 2D joint and 1D marginalized posterior probability distributions for $\xi$, $\Omega_m$, and $H_0$, obtained from the \textit{CMB}+\textit{BAO} (grey), \textit{CMB}+\textit{FS} (red), and \textit{CMB}+\textit{BAO}+\textit{FS} (blue) dataset combinations within the IDE model. The significant improvement in constraining power gained by the introduction of the \textit{FS} dataset is visually very clear. Note that $H_0$ is in units of ${\rm km}/{\rm s}/{\rm Mpc}$.}
\label{fig:triangularcmb}
\end{center}
\end{figure*}

\begin{figure*}
\begin{center}
\includegraphics[width=8cm]{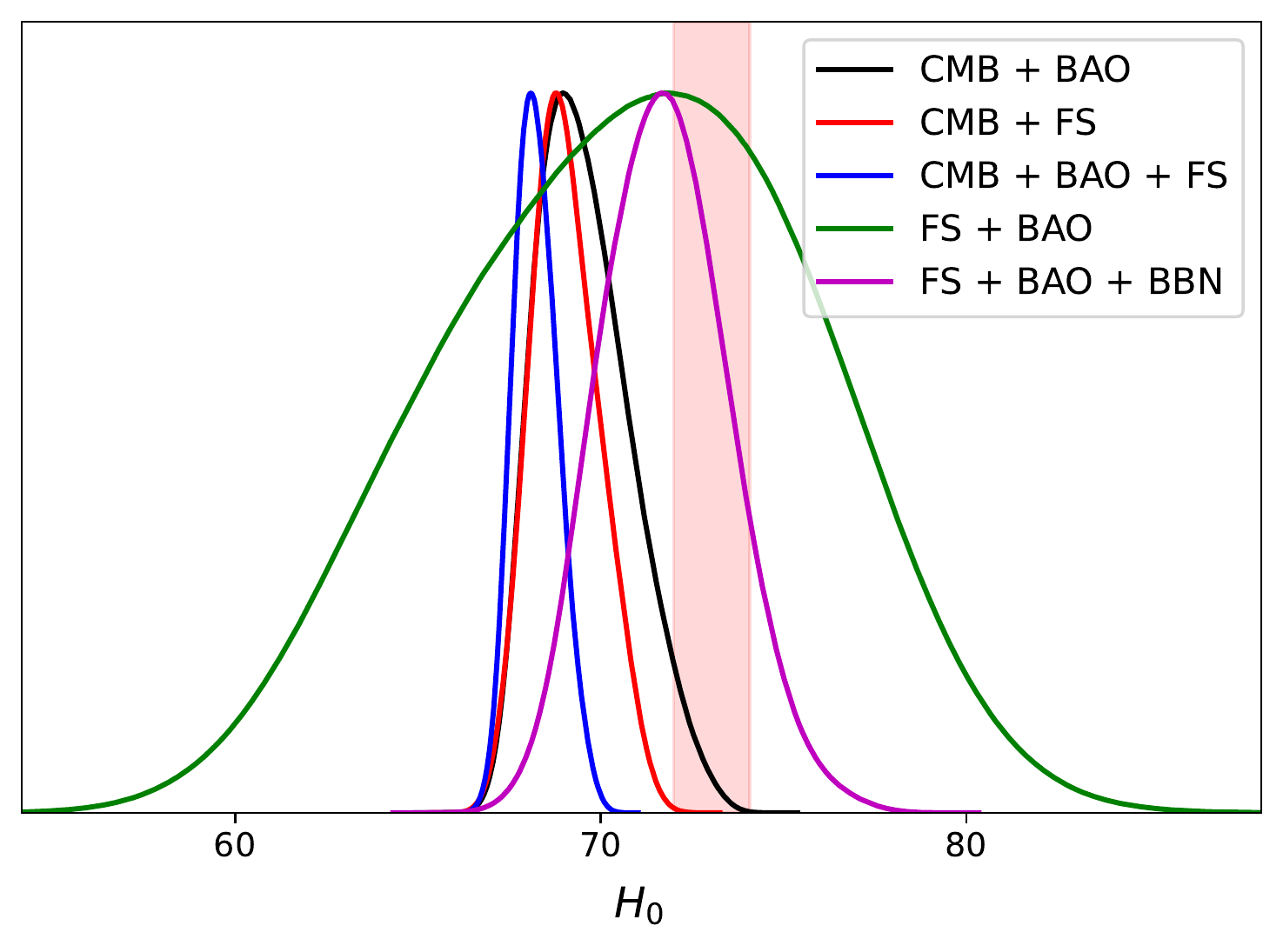} \,\,\,\,\,\,\,
\includegraphics[width=8cm]{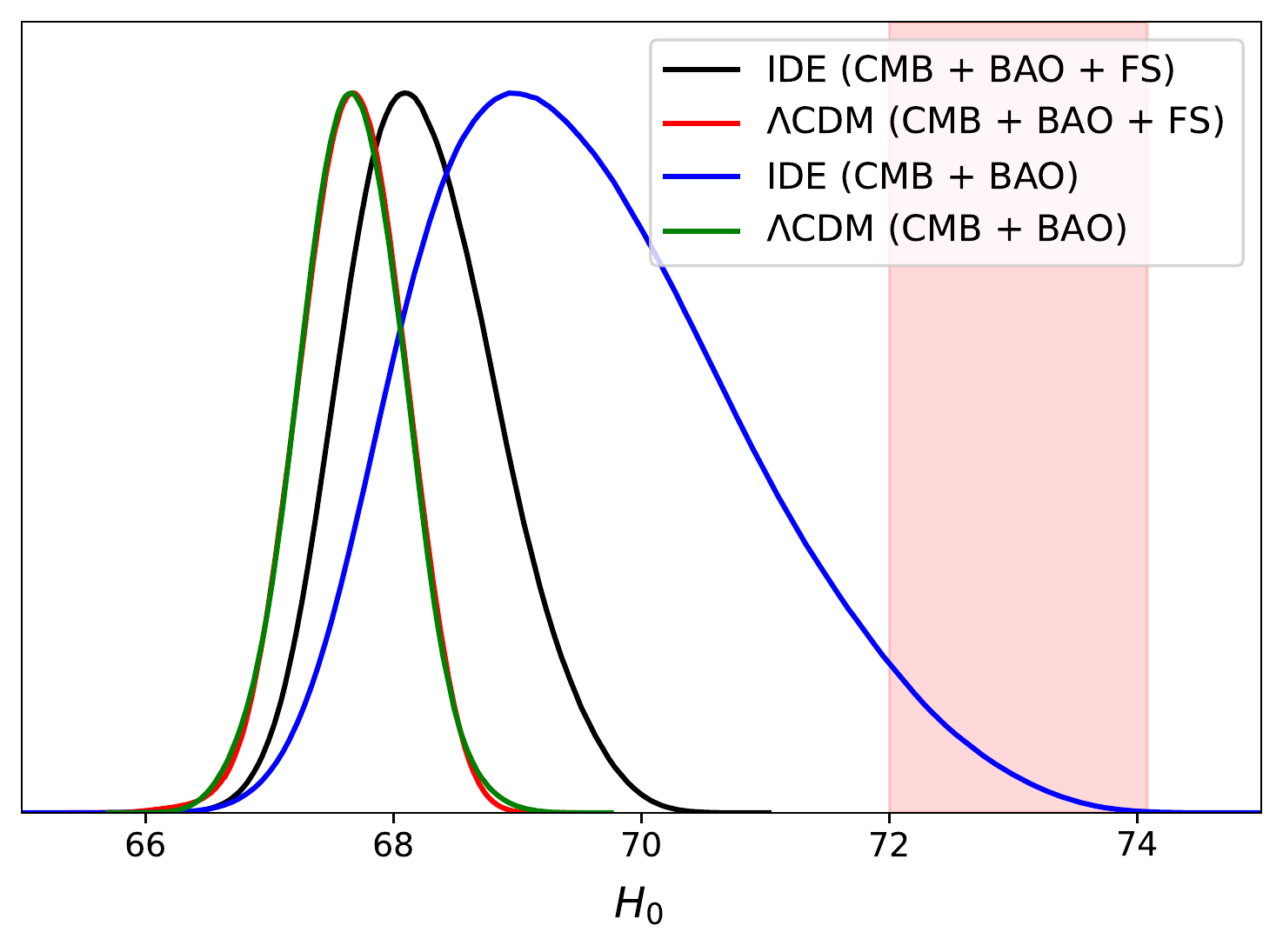} 
\caption{\textit{Left panel}: 1D marginalized posterior probability distributions for $H_0$, obtained from various dataset combinations (see the color coding) within the IDE model. The pink shaded band corresponds to the $1\sigma$ credible region for $H_0$ determined by the latest local distance ladder measurements constructed out of Cepheid-calibrated SNeIa~\cite{Riess:2021jrx}. \textit{Right panel}: 1D marginalized posterior probability distributions for $H_0$, obtained from the \textit{CMB}+\textit{BAO} and \textit{CMB}+\textit{BAO}+\textit{FS} dataset combination within the $\Lambda$CDM (green and red curves respectively) and IDE (blue and black curves respectively) models. From this figure it is clear that there is a significant gain in constraining power within the IDE model when including the \textit{FS} dataset, contrary to what is observed within $\Lambda$CDM. The pink shaded band is equivalent to that depicted in the left panel.}
\label{fig:marginalizedposteriorsh0}
\end{center}
\end{figure*}

\begin{figure*}
\begin{center}
\includegraphics[width=15cm]{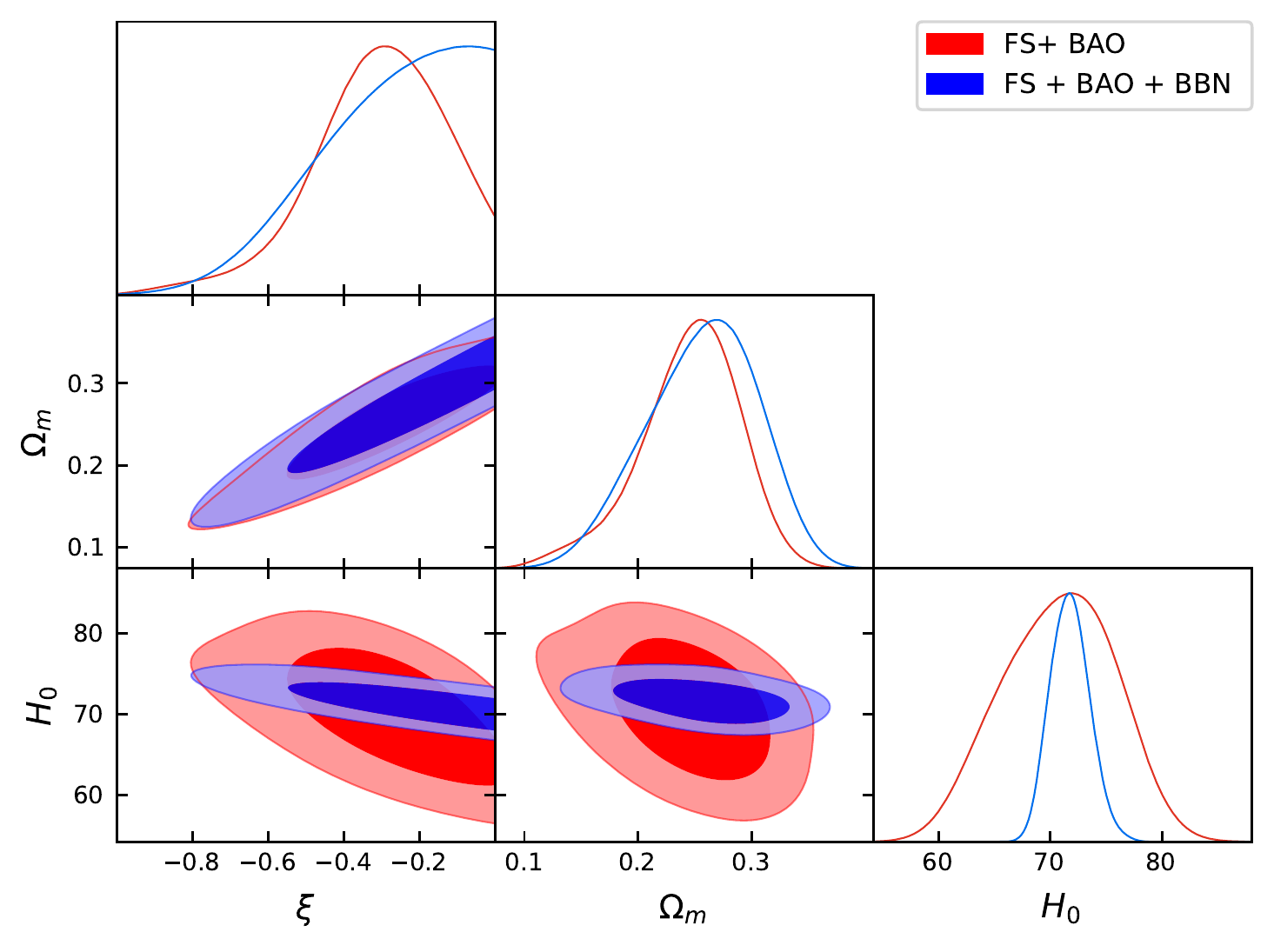} 
\caption{As Fig.~\ref{fig:triangularcmb}, but considering dataset combinations not involving the \textit{CMB} dataset: \textit{FS}+\textit{BAO} (red) and \textit{FS}+\textit{BAO}+\textit{BBN} (blue). It is clear that the inclusion of the \textit{BBN} prior on $\omega_b$ helps enormously with the determination of $H_0$, given that current galaxy clustering data mainly constrains $H_0$ through the sound horizon rather than through equality scale information (with the prior on $\omega_b$ sharpening the former), as discussed in the main text.}
\label{fig:triangularnocmb}
\end{center}
\end{figure*}

We begin by discussing the results obtained from dataset combinations involving the \textit{CMB} dataset: \textit{CMB}+\textit{BAO}, \textit{CMB}+\textit{FS}, and \textit{CMB}+\textit{BAO}+\textit{FS}. Afterwards, to better examine the role of shape versus geometrical information in setting constraints on IDE, we consider \textit{CMB}-free dataset combinations: \textit{FS}+\textit{BAO} and \textit{FS}+\textit{BAO}+\textit{BBN}. Our key results are reported in Tabs.~\ref{tab:parametersidecmb} and~\ref{tab:parametersidenocmb}, as well as Figs.~\ref{fig:triangularcmb},~\ref{fig:marginalizedposteriorsh0}, and~\ref{fig:triangularnocmb}. Note that for $\xi$ we shall always report the 95\% confidence level (C.L.) lower bound when the 68\%~C.L. interval is not consistent with a detection, whereas for all other cosmological parameters we report 68\%~C.L. intervals.

In the context of the \textit{CMB}+\textit{BAO} dataset combination, i.e the one driving state-of-the-art constraints on the DM-DE coupling strength $\xi$, we infer the lower bound $\xi>-0.48$ at 95\%~C.L. (while the 68\%~C.L. interval is $\xi=-0.22^{+0.18}_{-0.09}$). Note that this bound is slightly weaker than those reported for instance in Refs.~\cite{DiValentino:2019ffd,DiValentino:2019jae} because we here deliberately do not include BAO measurements from the 6dFGS and SDSS-MGS surveys: however, as will become clearer later, this particular \textit{CMB}+\textit{BAO} dataset combination allows for a cleaner assessment of the constraining power of shape and geometrical information in the IDE context. The reason is that our pipeline does not include the 6dFGS and SDSS-MGS FS measurements. Therefore, in this setting a full comparison of the constraining power of BAO measurements from BOSS DR12+6dFGS+SDSS-MGS versus FS measurements from the same combination is not possible. We thus make the choice of restricting our LSS measurements to the one galaxy survey for which we have both a BAO and FS pipeline readily available, i.e.,\ the BOSS DR12 galaxies. In the following, the limit $\xi>-0.48$ obtained from the \textit{CMB}+\textit{BAO} dataset combination will be the reference limit to which we will compare and assess any improvement brought about by the inclusion of the FS measurements.

Including FS measurements (\textit{CMB}+\textit{BAO}+\textit{FS}) by means of the joint BAO-FS EFTofLSS likelihood) tightens the limit on $\xi$ to $\xi>-0.12$ at 95\%~C.L., representing a significant improvement. There is, in fact, significant information contained in the BOSS DR12 FS measurements which helps disentangling IDE from $\Lambda$CDM. This is clear from the previous arguments in Sec.~\ref{subsec:ideeftoflss}, where we discussed the important role of the equality wavenumber in constraining IDE, as well as from Fig.~\ref{fig:residuals}.

To further examine the impact of the shape information, we now consider the \textit{CMB}+\textit{FS} dataset combination. The limit we obtain is $\xi>-0.35$ at 95\%~C.L., tighter than that obtained from the \textit{CMB}+\textit{BAO} dataset combination, implying that BOSS full-shape information carries more constraining power than the purely geometrical information from the reconstructed BAO wiggles after combining with CMB data. The inferred values of selected cosmological parameters from the three dataset combinations (\textit{CMB}+\textit{BAO}, \textit{CMB}+\textit{FS} and \textit{CMB}+\textit{BAO}+\textit{FS}) are reported in Tab.~\ref{tab:parametersidecmb}. In addition, the $\xi$-$\Omega_m$-$H_0$ triangular plot in Fig.~\ref{fig:triangularcmb} clearly shows the increased constraining power of the \textit{CMB}+\textit{FS} dataset combination with respect to the \textit{CMB}+\textit{BAO} one, as well as the overall significant improvement when considering the \textit{CMB}+\textit{FS}+\textit{BAO} dataset combination, which constitutes one of the novel results of this work.

At first glance, the above results might not appear to be surprising: after all, FS information encodes geometrical BAO information in the form of wiggles in $P(k)$, alongside other additional information contained in the broadband, non-wiggly part of $P(k)$. However, this conclusion should nonetheless surprise us, as a number of recent works have argued that shape and geometrical information carry similar constraining power in light of the precision of current data~\cite{Ivanov:2019hqk}. The EFTofLSS-based analysis of Ref.~\cite{Ivanov:2019hqk} argued that the comparable constraining power of shape and geometrical information in current BOSS data is purely a coincidence given the BOSS volume and redshift coverage, in combination with the efficiency of BAO reconstruction algorithms. However, Ref.~\cite{Ivanov:2019hqk} also argued that FS information is expected to dramatically supersede geometrical information in the context of future galaxy redshift surveys whose coverage will expand to significantly wider volumes and redshift ranges. Similar conclusions, albeit obtained from simpler (non-EFTofLSS) analyses of FS information, have been reached in at least two earlier works~\cite{Hamann:2010pw,Vagnozzi:2017ovm}. These works also discussed the possibility of such a conclusion being reversed in extensions to $\Lambda$CDM where shape information might play a crucial role, for instance in the context of beyond-$\Lambda$ DE models. Our results indeed provide a clear example in this sense: this further highlights the extremely important role of full-shape galaxy clustering information when probing non-standard extensions of $\Lambda$CDM, particularly in the DE sector, such as IDE models. In fact, as we have already argued earlier in Sec.~\ref{subsec:ideeftoflss}, information on the equality wavenumber $k_{\rm eq}$, not present in post-reconstruction BAO measurements, plays an important role in constraining IDE. A visual indication of the gain in constraining power when including the \textit{FS} dataset (in particular when moving from the \textit{CMB}+\textit{BAO} dataset combination to the \textit{CMB}+\textit{BAO}+\textit{FS} one) is given in the right panel of Fig.~\ref{fig:marginalizedposteriorsh0}, where it is clear that there is virtually no improvement within the $\Lambda$CDM scenario (green versus red curves), while the improvement is rather significant for the IDE model (blue versus black curves).

To further understand the role of geometrical and equality information in our results, we will closely follow the rationale of Refs.~\cite{Philcox:2020xbv,Farren:2021grl}, and consider additional dataset combinations not including CMB measurements. Full-shape measurements contain the imprint of two ``standard rulers'': the first is connected to the scale of the sound horizon at baryon drag $r_d$, and leaves its imprint in the position of the BAO wiggles, whereas the second is connected to the horizon wavenumber at matter-radiation equality $k_{\rm eq}$, and governs the position of the turnaround in the power spectrum, see Sec.~\ref{subsec:ideeftoflss}. Information from both scales can be used to constrain $H_0$ or, for that matter, any other parameter which affects $r_d$ and $k_{\rm eq}$ either directly or indirectly.\footnote{An example of a parameter indirectly influencing $r_d$ or $k_{\rm eq}$ is one that does so through the induced shifts in other parameters. For instance, besides their direct impact on $\rho_c(z)$ clear from Eq.~(\ref{eq:solutionrhoc}), which obviously affects $z_{\rm eq}$, IDE models typically lead to shifts in $\omega_c$, which translate into additional shifts in $r_d$ and in the redshift of matter-radiation equality $z_{\rm eq}$, and therefore in $k_{\rm eq}$.}

One way of assessing the extent to which constraints involving FS data are driven by geometrical BAO information is to artificially limit or even remove any information on $r_d$ whatsoever. To zeroth order, this can be achieved by removing any external calibration on $r_d$, which is tantamount to removing any external informative prior on the physical baryon density $\omega_b$: FS analyses not including CMB data typically include a BBN prior on $\omega_b$ (see e.g.\ Ref.~\cite{Philcox:2020vvt}), given that knowledge of $\omega_b$ (which controls the pre-recombination baryon sound speed) is required to compute $r_d$ and therefore calibrate the geometrical BAO information. Notice that this procedure is not exactly equivalent to marginalizing over $r_d$, which is the proper procedure to follow to remove any geometrical information and extract only information from the broadband part of FS measurements. Nevertheless, given the precision of current galaxy clustering data, this procedure works (as argued in Ref.~\cite{Philcox:2020xbv}), whereas for future, more precise FS data, the BAO features and the small-scale $\omega_b$-induced Jeans suppression in FS measurements could in principle cross-calibrate each other, effectively generating an $\omega_b$ or $r_d$ prior. The later work of Ref.~\cite{Farren:2021grl} constructs an improved method to remove geometrical information in galaxy clustering measurements in future surveys. For the purposes of the present work however, given the precision of BOSS FS measurements, removing any external informative prior on $\omega_b$ is sufficient to ensure that the sound horizon information is essentially completely (or mostly) removed, and we will therefore follow the methodology of Ref.~\cite{Philcox:2020xbv}.

In light of the above discussion, we shall now consider the \textit{FS}+\textit{BAO}+\textit{BBN} and \textit{FS}+\textit{BAO} dataset combinations: the former allows for the exploitation of sound horizon information through the BBN prior on $\omega_b$, which calibrates $r_d$, whereas such information is absent in the latter.\footnote{Note that any dataset combination involving the \textit{CMB} dataset contains information on $\omega_b$ and therefore $r_d$, as $\omega_b$ is determined by the relative height of the odd versus even peaks.} When considering \textit{FS}+\textit{BAO}+\textit{BBN} we infer $\xi>-0.62$ at 95\%~C.L., whereas removing the BBN prior on $\omega_b$ (thus considering the \textit{FS}+\textit{BAO} dataset combination) surprisingly returns exactly the same lower limit on $\xi$, although in this case the 68\%~C.L. interval is consistent with a $\simeq 1.5\sigma$ detection: $\xi=-0.31^{+0.20}_{-0.13}$. This further confirms the fact that the constraints on $\xi$ are driven to an important amount by broadband information rather than by just geometrical information: if the latter were true, the constraints on $\xi$ should have become significantly weaker when removing the \textit{BBN} dataset, contrary to what we observe. The inferred values of selected cosmological parameters from the two CMB-free dataset combinations discussed above are reported in Tab.~\ref{tab:parametersidenocmb}, whereas in Fig.~\ref{fig:triangularnocmb} we present the $\xi$-$\Omega_m$-$H_0$ triangular plot obtained from these same two dataset combinations. 

It is important to explore the impact of the inclusion of FS measurements on the inference of parameters other than the coupling $\xi$. There are two key parameters in IDE cosmologies: the matter density parameter $\Omega_m$ and the Hubble constant $H_0$. The correlation between $\xi$ and $\Omega_m$ can be understood in terms of the direction of energy flow between DM and DE. Since $\xi<0$, energy flows from DM to DE, and therefore the DM density today (and correspondingly $\Omega_m$) will be smaller~\cite{DiValentino:2019ffd,DiValentino:2019jae}. This is also clear from Eq.~(\ref{eq:solutionrhoc}), where the evolution of $\rho_c$ carries an extra term proportional to $\xi$ in addition to the standard $\Lambda$CDM term $\rho_c \propto (1+z)^3$. However, as $\Omega_mh^2$ is tightly constrained by CMB data, the decrease in $\Omega_m$ requires an increase in $H_0$. This explains the mutual $\xi$-$\Omega_m$-$H_0$ degeneracies we observe: a direct correlation between $\xi$ and $\Omega_m$, and an inverse correlation between $\xi$ and $H_0$, with a significantly weaker inverse correlation between $\Omega_m$ and $H_0$. Focusing on $\Omega_m$ and $H_0$, we find that their determination is significantly improved by the inclusion of the \textit{FS} dataset on top of the \textit{CMB}+\textit{BAO} dataset combination. In particular, the uncertainties on $\Omega_m$ and $H_0$ are decreased by factors of $\sim$3 and $\sim$2 respectively. This improvement (particularly for $\Omega_m$) is due to the important role of the equality scale in FS measurements. In fact, equality information in FS measurements directly constrains the so-called ``shape parameter'' $\Gamma \equiv \Omega_mh$~\cite{Efstathiou:1992sy,2dFGRS:2001ybp}. In combination with CMB measurements constraining $\Omega_mh^2$ through the early integrated Sachs-Wolfe~\cite{Hou:2011ec,Cabass:2015xfa,Kable:2020hcw,Vagnozzi:2021gjh} and lensing effects~\cite{Baxter:2020qlr}, and geometrical BAO information, this helps pinning down $\Omega_m$, particularly in the context of extended models (see for instance the discussion in Section~2 of Ref.~\cite{Vagnozzi:2020rcz} in the context of spatial curvature). In Sec.~\ref{subsec:ideeftoflss} we already emphasized the important role of broadband information (and in particular information related to $k_{\rm eq}$) in the context of IDE models, which is anyhow clear from the monopole curves in Fig.~\ref{fig:residuals}.

Moving to the CMB-free dataset combinations discussed earlier (\textit{FS}+\textit{BAO} and \textit{FS}+\textit{BAO}+\textit{BBN}), we note that the parameter most significantly affected by the addition of the BBN prior on $\omega_b$ is the Hubble constant $H_0$, whose uncertainty decreases by a factor of $\sim 3$ once the \textit{BBN} dataset is included. This is not surprising, given that current galaxy clustering data mostly constrain $H_0$ through the sound horizon rather than from equality scale information~\cite{Philcox:2020xbv}. The uncertainty we infer on $H_0$ of order $5\,{\rm km}/{\rm s}/{\rm Mpc}$ is in fact comparable to that obtained in Ref.~\cite{Philcox:2020xbv} for the $\Lambda$CDM model (of course, our uncertainties are slightly larger because of the extra parameter $\xi$ and its correlation with $H_0$). To conclude this section, it is clear that full-shape information is highly precious when dealing with non-trivial extensions of the $\Lambda$CDM model, particularly within IDE cosmologies. The addition of FS information further limits the available parameter space for the DM-DE coupling strength, even in the absence of CMB data. 

\section{Conclusions}
\label{sec:conclusions}

Driven by important theoretical advances, in the last couple of years significant efforts have gone into extracting Large-Scale Structure (LSS) clustering information beyond geometrical information contained in the reconstructed BAO peaks, by exploiting the information content of the full-shape (FS) power spectrum of LSS tracers. The power of FS measurements and their ability to not only sharpen constraints on the parameters of the baseline $\Lambda$CDM model, but perhaps more importantly test extensions thereof, is by now established. In the wake of these important developments, the present work is the first to test interacting dark energy (IDE) cosmologies in light of state-of-the-art redshift-space galaxy clustering data, with a robust theoretical modeling of the underlying mildly non-linear power spectrum.

IDE cosmologies feature non-gravitational interactions between dark matter (DM) and dark energy (DE) and have experienced nothing short of a resurgence in recent years. In the context of our work, IDE models provide a unique phenomenological stage to test the power and advantages of certain cosmological observations with respect to others. Specifically, our goal is to assess the improvement gained by considering FS measurements when constraining the DM-DE coupling strength $\xi$, taking as baseline state-of-the-art constraints which only consider geometrical BAO measurements (see e.g.\ Ref.~\cite{DiValentino:2019jae}). Focusing on an IDE model [see Eq.~(\ref{eq:Q})] which is well-studied, mostly due to its simplicity and connection to coupled quintessence, we have demonstrated how the inclusion of FS measurements significantly sharpens constraints on $\xi$, which is now safely constrained to $\vert \xi \vert \lesssim 0.1$ for the model considered.

Our limits on $\xi$ imply that DM and DE cannot exchange energy at a rate greater than ${\cal O}(10^{-48})\,{\rm g}/{\rm s}/{\rm cm}^3$, or equivalently that they cannot exchange more than $\approx 7\%$ of the critical energy density over the course of a Hubble time, a quantity which is way too small to have any bearings on the coincidence problem (as was already known). Moreover, these tight constraints further limit the ability of IDE cosmologies to play an important role in the context of the Hubble tension. Crucially, the same limits also constrain the DE effective equation of state (EoS) $w_{x,{\rm eff}}=w_x+\xi/3 \gtrsim -1.04$, consistent with independent constraints on the DE EoS from various probes, which strongly limit the ability to move deep into the phantom region (see e.g. Refs.~\cite{Zhao:2017cud,Vagnozzi:2018jhn,Gerardi:2019obr,Vagnozzi:2019ezj,Visinelli:2019qqu,Bonilla:2020wbn,Colgain:2021pmf,Teng:2021cvy,Raveri:2021dbu,Sharma:2022ifr}).~\footnote{Our results, and particularly the inferred values of $H_0$, can also be compared to those of Ref.~\cite{Krishnan:2021dyb}, given the background equivalence between IDE and dynamical DE models.}

Our results highlight the extremely important role of FS measurements in the era of precision cosmology, particularly in further sharpening constraints on models beyond $\Lambda$CDM: for instance, a recently well-documented example in this context is that of the early dark energy model~\cite{Hill:2020osr,Ivanov:2020ril,DAmico:2020ods,Niedermann:2020dwg,Smith:2020rxx,Ye:2021nej,Poulin:2021bjr,LaPosta:2021pgm,Smith:2022hwi,Jiang:2022uyg}.~\footnote{See also Ref.~\cite{Rudelius:2022gyu} for related discussions on theoretical difficulties early dark energy faces.} Furthermore, contrary to what occurs within the $\Lambda$CDM model and simple extensions thereof, we have demonstrated that constraints obtained by combining CMB and FS data are tighter than those obtained by combining CMB and BAO measurements, suggesting that in the context of IDE models there is significant information contained in the broadband of the power spectrum. We have extensively tested these findings, discussing the important role played by the equality scale $k_{\rm eq}$ in constraining IDE cosmologies.

This work is the first to robustly test DM-DE interactions from state-of-the-art redshift-space galaxy clustering measurements, and the constraints we report on the coupling strength $\xi$ are among the strongest ever presented in the literature. At the same time, there are many interesting follow-up directions one could envisage. While here we have considered FS measurements from the BOSS DR12 sample, it would be very interesting to repeat a similar analysis using the much more recent eBOSS DR16 measurements~\cite{eBOSS:2020yzd}. On the other hand, we expect the role of FS galaxy clustering information to become even more important in this context with the advent of future LSS surveys, such as Euclid~\cite{Amendola:2016saw} and DESI~\cite{DESI:2016fyo}. It would be interesting to forecast the ability of these surveys to constrain DM-DE interactions when combined with future CMB data, for instance from \textit{Simons Observatory}~\cite{SimonsObservatory:2018koc,SimonsObservatory:2019qwx} or CMB-S4~\cite{CMB-S4:2016ple}. Finally, it would be very interesting to test other well-motivated beyond-$\Lambda$CDM models in light of the same FS measurements adopted here. Most of these ideas are the subject of work in progress, on which we hope to report soon.

\begin{acknowledgments}
\noindent S.V. thanks Misha Ivanov for useful discussions. R.C.N. acknowledges financial support from the Funda\c{c}\~{a}o de Amparo \`{a} Pesquisa do Estado de S\~{a}o Paulo (FAPESP, S\~{a}o Paulo Research Foundation) under the project No.~2018/18036-5. S.V. is supported by the Isaac Newton Trust and the Kavli Foundation through a Newton-Kavli Fellowship, and by a grant from the Foundation Blanceflor Boncompagni Ludovisi, n\'{e}e Bildt. S.V. acknowledges a College Research Associateship at Homerton College, University of Cambridge. S.K. gratefully acknowledges support from the Science and Engineering Research Board (SERB), Govt. of India (File No.~CRG/2021/004658). E.D.V. is supported by a Royal Society Dorothy Hodgkin Research Fellowship. O.M. is supported by the Spanish grants PID2020-113644GB-I00, PROMETEO/2019/083 and by the European ITN project HIDDeN (H2020-MSCA-ITN-2019//860881-HIDDeN).

\end{acknowledgments}

\bibliography{IDEFS}

\end{document}